\documentclass{aastex63}

\newcommand{\dtb}{$\Delta T_\mathrm{b}$}
\newcommand{\lo}{\textsf{\textbf{x}}}
\newcommand{\hi}{\textsf{\textbf{o}}}

\newcommand{\jastp}{\textit{J. Atmos. Sol-Terr. Phys.}}

\received{2020 May 17}
\accepted{2020 September 7}
\submitjournal{ApJ}
\shorttitle{Submillimeter ARs and the Solar Cycle}
\shortauthors{Gim{\'e}nez de Castro, C.G., et al.}

\begin{document}

\title{The Submillimeter Active Region Excess Brightness Temperature during Solar Cycles 23 and 24}

\correspondingauthor{C. Guillermo Gim{\'e}nez de Castro}
\email{guigue@craam.mackenzie.br, guigue@gcastro.net}

\author[0000-0002-8979-3582]{C. Guillermo Gim{\'e}nez de Castro}
\affiliation{Centro de R{\'a}dio Astronomia e Astrof{\'\i}sica Mackenzie \\
  Rua da Consolação, 896\\ S{\~a}o Paulo, 01302-907, Brazil}
\affiliation{Instituto de Astronom{\'\i}a y F{\'\i}sica del Espacio, CONICET-UBA,
  Ciudad Universitaria\\ Buenos Aires, Argentina}

\author{Andr{\'e} L. G. Pereira}
\affiliation{Centro de R{\'a}dio Astronomia e Astrof{\'\i}sica Mackenzie \\
  Rua da Consolação, 896\\ S{\~a}o Paulo, 01302-907, Brazil}

\author[0000-0001-7144-7967]{J. Fernando Valle Silva}
\affiliation{Centro de R{\'a}dio Astronomia e Astrof{\'\i}sica Mackenzie \\
  Rua da Consolação, 896\\ S{\~a}o Paulo, 01302-907, Brazil}

\author[0000-0002-5897-5236]{Caius L. Selhorst}
\affiliation{N{\'u}cleo de Astrof{\'\i}sica, Universidade Cruzeiro
  do Sul / Universidade Cidade de S{\~a}o Paulo, S{\~a}o Paulo, SP,
  Brazil}

\author[0000-0001-9311-678X]{Cristina H. Mandrini}
\affiliation{Instituto de Astronom{\'\i}a y F{\'\i}sica del Espacio, CONICET-UBA,
  Ciudad Universitaria\\ Buenos Aires, Argentina}
\affiliation{Universidad de Buenos Aires, Facultad de Ciencias Exactas y Naturales, 1428 Buenos Aires, Argentina}

\author{Germ{\'a}n D. Cristiani}
\affiliation{Instituto de Astronom{\'\i}a y F{\'\i}sica del Espacio, CONICET-UBA, 
  Ciudad Universitaria\\ Buenos Aires, Argentina}
\affiliation{Universidad de Buenos Aires, Facultad de Ciencias Exactas y Naturales, 1428 Buenos Aires, Argentina}

\author[0000-0002-7501-3231]{Jean-Pierre Raulin}
\affiliation{Centro de R{\'a}dio Astronomia e Astrof{\'\i}sica Mackenzie \\
  Rua da Consolação, 896\\ S{\~a}o Paulo, 01302-907, Brazil}

\author[0000-0002-1671-8370]{Adriana Valio}
\affiliation{Centro de R{\'a}dio Astronomia e Astrof{\'\i}sica Mackenzie \\
  Rua da Consolação, 896\\ S{\~a}o Paulo, 01302-907, Brazil}

\begin{abstract}
  We report the temporal evolution of the excess brightness temperature   \dtb\ above solar active regions (ARs) observed with the Solar   Submillimeter Telescope (SST) at 212 ($\lambda=1.4$~mm) and 405 GHz   ($\lambda=0.7$~mm) during Cycles 23 and 24. Comparison with the   sunspot number (SSN) yields a Pearson's correlation coefficient $R=0.88$ and   $0.74$ for 212 and 405 GHz, respectively. Moreover, when only Cycle   24 is taken into account the correlation coefficients go to 0.93 and 0.81   for each frequency. We derive the spectral index $\alpha$ between SST frequencies and found a slight anti-correlation with the SSN ($R = -0.25$); however, since the amplitude of the variation is lower than the  standard deviation we cannot draw a definite conclusion. Indeed, $\alpha$ remains almost constant within the uncertainties   with a median value of $\approx 0$ characteristic of an optically thick   thermal source. Since the origin of the AR submillimeter radiation   is thermal continuum produced at chromospheric heights, the strong   correlation between \dtb\ and the magnetic cycle evolution could be related to the available free magnetic energy to be released in reconnection events. 
\end{abstract}


\section{Introduction}
\label{sec:Intro}

Solar active regions (ARs) are the \textit{`` The totality of all observable phenomena preceding, accompanying and following the birth of sunspots including radio-, X-, EUV- and particle emission.''} \citep[extracted from ][]{vanDrielGreen:2015}. In particular at radio frequencies they are areas of enhanced brightness on the solar disk observed near sunspots. At microwaves \cite{Selhorstetal:2014} have shown the correlation between the monthly average of the maximum and mean excess brightness temperatures at 17~GHz of ARs and the solar cycle, this work is based on images obtained by the Nobeyama Radioheliograph \citep[NoRH,][]{Nakajima:1994} with synthesized beam sizes of the order of 10~arcsec. \cite{Kundu:1970}  conducted the first millimeter observations using the 11~m (36-foot) single dish antenna installed at Kitt Peak (USA).  Around 40 solar maps were obtained at 33.3, 85.6 and 249.8 GHz (although only three at the last frequency) with spatial resolutions between $3^\prime.5$ and $1^\prime.2$ from which he derived typical brightness temperatures of 1000, 700 and 150~K for the three frequencies, respectively. Using the Caltech Submillimeter Observatory (CSO) \cite{Bastianetal:1993} were able to analyze brightness observed at 352~GHz, associated with H$\alpha$ filaments and determined temperatures close to or below that of the quiet Sun. Moreover, they observed a prominence at 352 and 240~GHz; in both cases they arrived to the conclusion that the millimeter/submillimeter emission of filaments and prominences is optically thin. At higher frequencies \cite{LindseyKopp:1995} used the James Clerck Maxwell Telescope (JCMT) to make maps of the Sun at  250, 350 and 855~GHz with beam sizes between 16 to 5~arcsec. They showed for the first time that sunspots, when observed at submillimeter wavelengths, have an umbra cooler than the quiet Sun, while their penumbrae could have temperatures similar to that of the quiet Sun; furthermore, in some ARs, sunspots could be as hot as the surrounding plages. Indeed, \cite{IwaiShimojo:2015} used the 45~m Nobeyama telescope to determine the temperature of the umbra and penumbra at 85 and 115~GHz and also reported finding an umbra  cooler than the quiet Sun. These findings were recently confirmed by \cite{Loukitchevaetal:2017} and \cite{Iwaietal:2017}, using the images obtained with the Atacama Large Millimeter Array (ALMA) with a synthesized beam of 0.5~arcsec at frequencies of 100 ($\lambda=3.0$~mm) and 239~GHz ($\lambda=1.2$~mm). With lower spatial resolution, \cite{Silvaetal:2005} analyzed ARs during 23 days at 212 ($\lambda=1.4$~mm) and 405~GHz ($\lambda=0.7$~mm) with the Solar Submillimeter Telescope, adding complementary data from NoRH at 17 and 34~GHz ($\lambda=8.8$~mm). They concluded that an overall AR behaves as an optically thick thermal source, because of the average flux density spectral index $\alpha\simeq 2$. This result was confirmed by \cite{ValleSilvaetal:2020}, who, besides SST and NoRH, included single dish ALMA maps at 107 ($\lambda=2.8$~mm) and 239~GHz ($\lambda=1.2$~mm). Moreover, they showed that an AR without associated sunspots has a harder flux density spectral index.\\

However, a study comprising a whole solar cycle at submillimeter wavelengths, analyzing possible changes in the AR characteristics, was never before attempted. \cite{GarciaPereiraetal:2020} used SST and NoRH observations obtained between 2001 and 2017, and an artificial intelligence (AI) algorithm, based on neural networks and computer vision, to identify, extract, classify and obtain physical properties of ARs observed in more than 16,000 maps. They showed that the number of ARs per year follows the solar cycle, and confirmed, at a solar cycle time-scale, \cite{Silvaetal:2005} and \cite{ValleSilvaetal:2020} conclusions.  In the present work, we use the statistical results from \cite{GarciaPereiraetal:2020} and focus in the temporal evolution of the physical properties of ARs at submillimeter wavelengths.

\section{Data Reduction and Analysis}
\label{sec:method}

\begin{figure}
\includegraphics[width=0.33\textwidth]{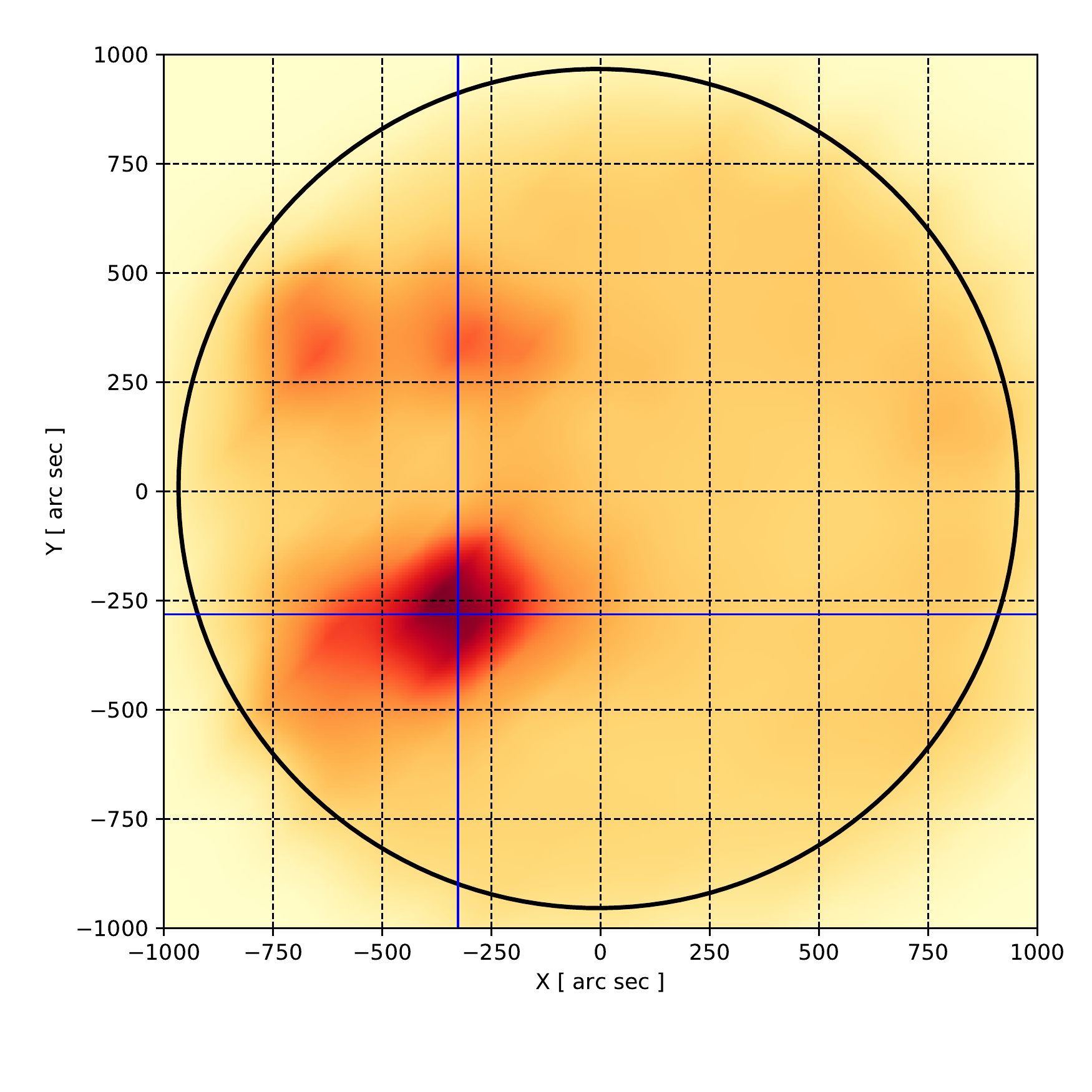}  
\includegraphics[width=0.33\textwidth]{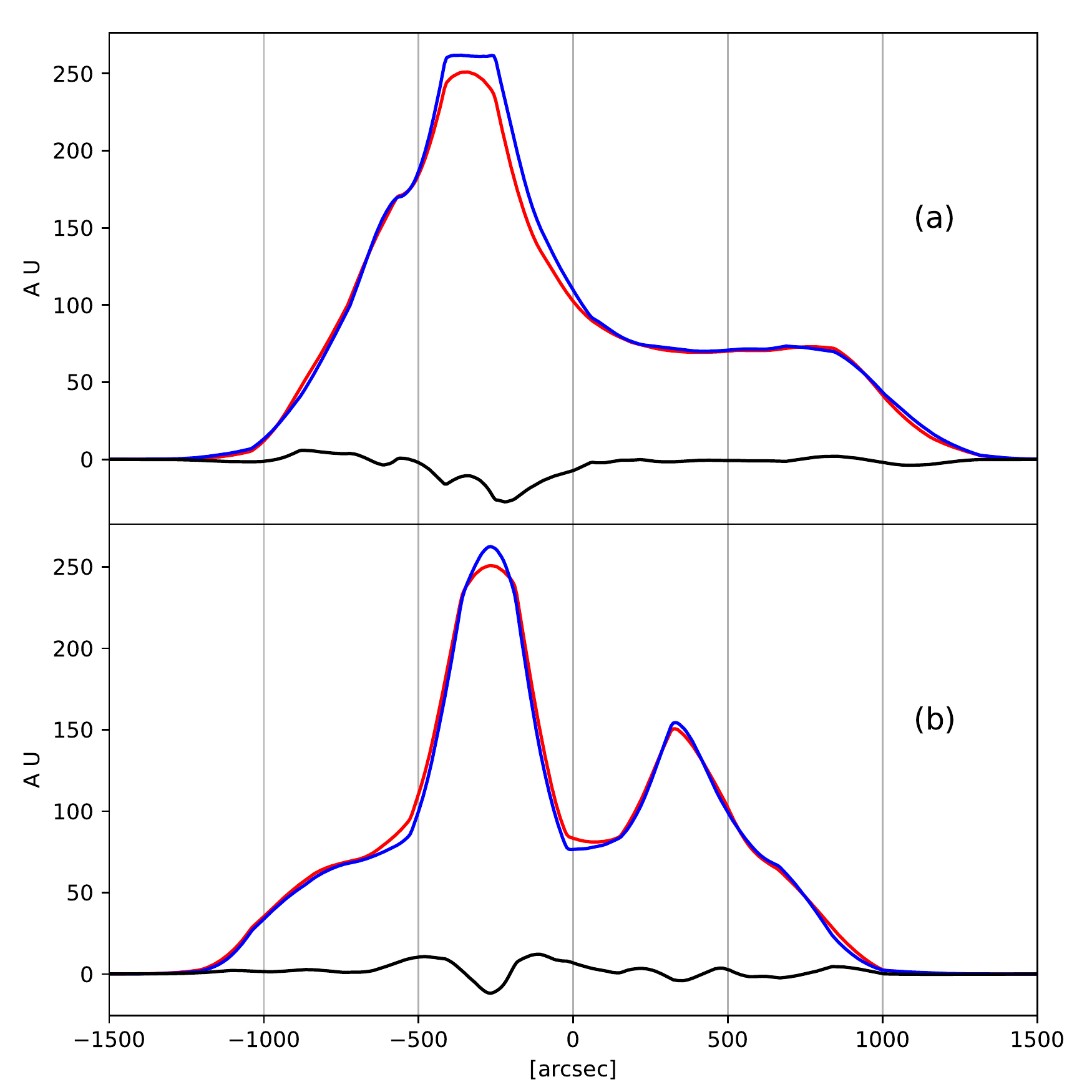}  
\includegraphics[width=0.33\textwidth]{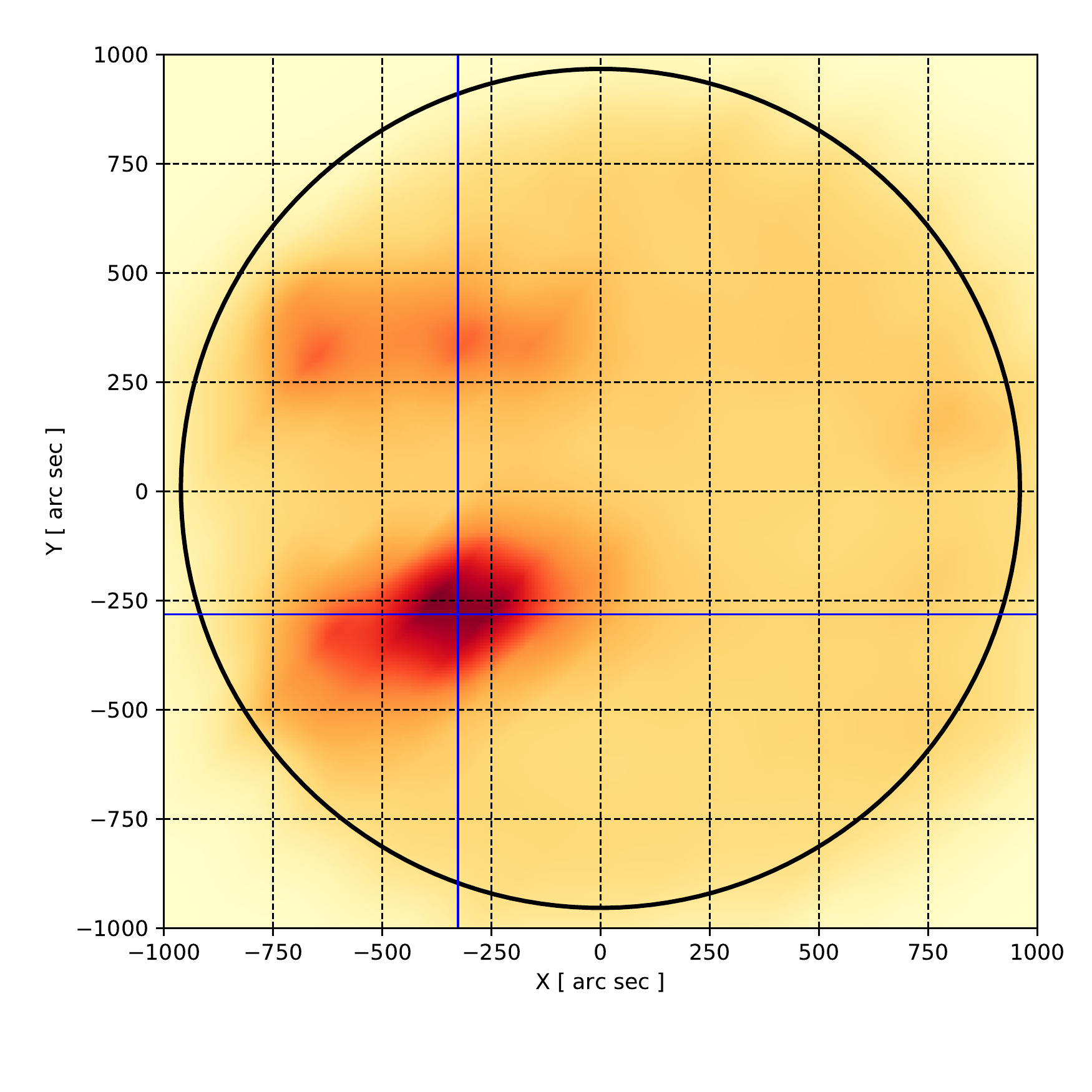}  

\includegraphics[width=0.33\textwidth]{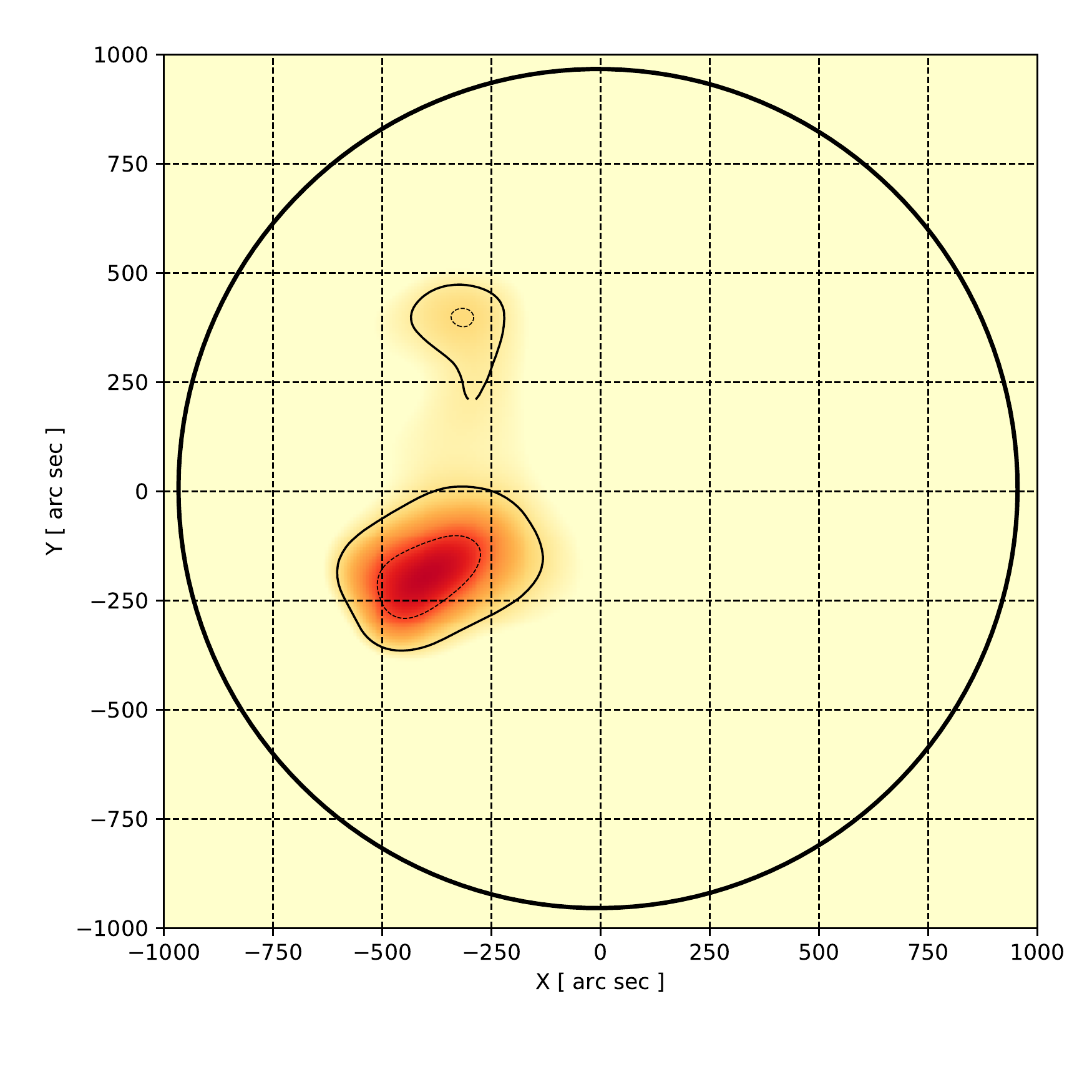}  
\includegraphics[width=0.33\textwidth]{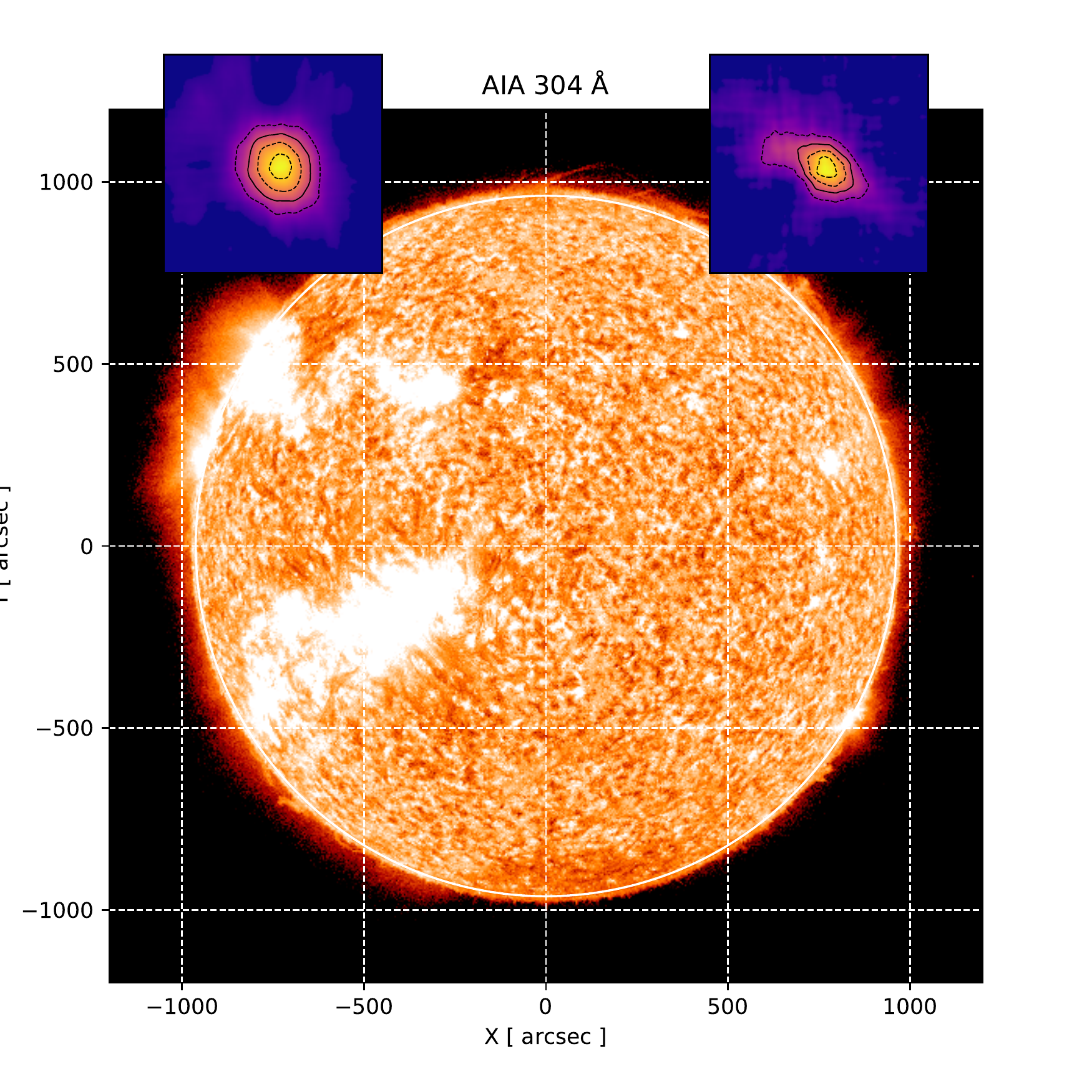}  
\includegraphics[width=0.33\textwidth]{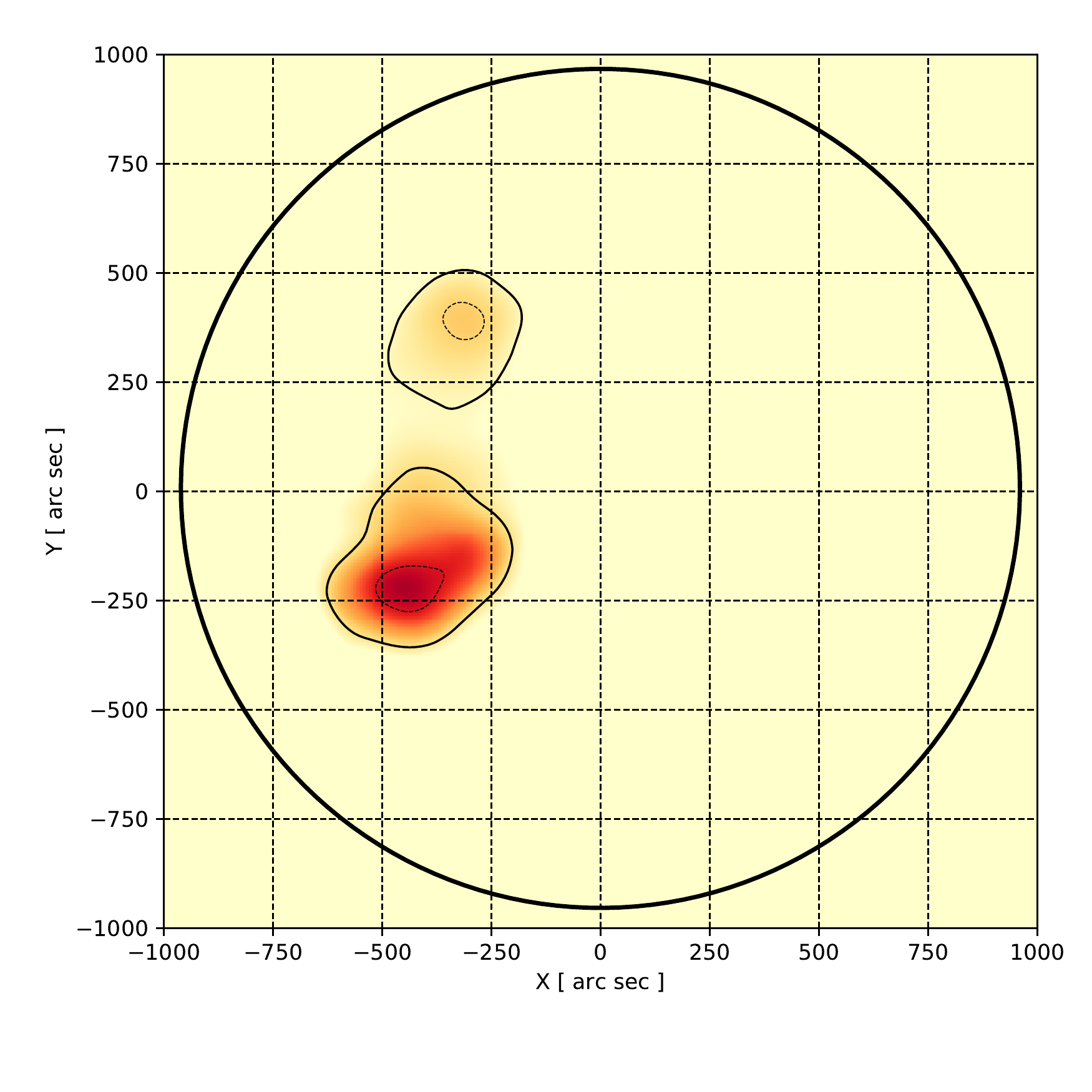}  
\caption{\textit{Top panel, left}: a simulated 212 GHz map, blue straight lines represent cuts where intensity profiles were determined. \textit{Middle}: a) red (blue) is the profile across a horizontal cut over the disk at 212 GHz (405 GHz); b) the same as a) for a vertical cut. The black curve represents the difference between profiles. \textit{Right}: a simulated 405 GHz map, blue straight lines represent the place where intensity profiles were determined. \textit{Bottom panel, left}: an observed solar map at 212 GHz taken on 2011-03-26 at 15:00 UT. \textit{Middle}: an SDO/AIA 304~\AA\ image taken on 2011-03-26 at 15:02 UT, the insets show the 212 (left) and 405 (right) GHz beams. \textit{Right}: an observed solar map at 405 GHz taken on 2011-03-26 at 15:00 UT.}
\label{fig:mapas}
\end{figure}

The SST \citep{Kaufmannetal:2008} is a multi-beam radio telescope,
installed at the Complejo Astronómico El Leoncito (CASLEO, Argentina),
composed of six independent radiometers arranged in the focus of a
Cassegrain 1.5~m antenna that observe simultaneously. Daily and
continuous observations started on March 2001 and, with few stops for maintenance work, SST already gathered almost 20 years of data. SST main objective is to spectrally and spatially characterize flares at the frequencies of 212 and 405~GHz. SST nominal half-power beam widths (HPBW) are 4 and 2 arcmin for 212 and 405~GHz, respectively. However, structural problems of the antenna produce deformations to the beam shapes, more pronounced at 405 GHz. Indeed, the high frequency beams are elongated with sizes $\sim 4^\prime$ and $\sim 2^\prime$ at the -3\ dB (50~\%) level along the \textit{major} and \textit{minor axis}, respectively. By making radial solar maps and applying a tomographic reconstruction method, we determine the SST beams up to the level of -13~dB \citep[5~\%, ][]{Costaetal:2002}. The reconstructed beams have similar characteristics among receivers of equal frequency. In order to assess the effect of the beam shapes on the observed sources, we have simulated observations convolving high resolution EUV images with the SST reconstructed beams. Simulations, instead of observations, are free from other aspects that affect differently observations at the two frequencies, such as atmospheric attenuation, receivers temperatures and gains. On the other hand, we choose an SDO/AIA 304~\AA\ image \citep{Lemenetal:2012} because it is representative of thermal emission from the chromosphere/transition region, has a high contrast with respect  to the quiet Sun and presents features of different sizes. Figure \ref{fig:mapas} shows the convolution of an SDO/AIA 304 \AA\ image  taken on 26 March 2011 at 15:02 UT with a pixel size of $0.6\arcsec \times 0.6\arcsec$  with a 212 GHz (top left) and  405 GHz beam (top right). Despite the sources have different sizes, all of them   present similar characteristics after the convolution with the beams at both wavelengths. This can be verified in the profiles also shown in the same figure (top middle): (a) is a profile in the X direction and (b) in the Y direction, their positions are represented with blue straight lines in the simulated maps. Blue and red curves are the profiles across the disk in function of position for the 212 and 405 GHz synthesized  maps, respectively. Black curves are the difference between profiles. The \textit{real} observed maps are shown in the bottom panel: the 212 GHz map on the left and the 405 GHz map on the right. In the middle panel we include the SDO/AIA image used to synthesize the SST maps, with the antenna beams depicted in the insets. \\

The uncooled receivers have typical temperatures $T_R\approx 2500 - 3500$~K, similar bandwidths $\Delta\nu=4$~GHz, and integration times $\delta t = 40$~milliseconds. During a solar observation the maximum expected antenna temperature is $T_a \approx 6000$~K (i.e. no atmospheric attenuation and beam efficiency $\approx 1$), yielding  a root mean square system temperature
$$\Delta T_\mathrm{system} = \frac{T_R+T_a}{\sqrt{\delta t \ \Delta\nu}} \lessapprox 1\ \mathrm{K} \ .$$ 
More important than the noise temperature is the atmospheric opacity at El Leoncito which can mildly attenuate the radiation at 212~GHz but can severely reduce the emission at 405~GHz \citep{Meloetal:2005, Cassianoetal:2018}.\\

SST provides several solar maps per day as a calibration procedure. These maps are the basic data of our work. SST solar maps are typically made out of azimuth raster scans (\textit{on-the-fly maps}) of 60 arcmin length.  The elevation changes at every scan to make a square that covers the entire solar disk. A typical map has a separation between scans equal to 2~arcmin. The scan speed is between $0.1 \ \mathrm{to}\ 0.2 \ \mathrm{deg \ s^{-1}}$. Since we use 40 millisecond integrated data, these speeds produce a separation between successive points in azimuth between 0.24 to 0.48 arcmin. Therefore, observations are oversampled in azimuth creating a rectanglular pixel of $(0.24 - 0.48) \times 2\ \mathrm{arcmin}^2$; maps have either $375 \times 31$ pixels or $187 \times 31$ pixels.  We interpolate the raw data to create a square map with a field of view of 31~arcmin and $288 \times 288$ pixels.  Then, the map is rotated with north up and the collected intensity units are converted to temperature, using the Sun brightness temperature as a reference \citep[see
][]{Silvaetal:2005}. We note that there is no need to equalize the beam sizes by convolving 405~GHz observations with the 212~GHz beam and vice versa, since, from the results presented above, both beams have similar spatial response.\\

The artificial intelligence (AI) procedure implemented to detect ARs in SST maps is based on an artificial convolutional neural network \citep[ConvNet,][]{Ciresanetal:2011,OSheaNash:2015} composed of four convolutional layers and two fully connected layers. Every convolutional layer has one max pooling and one flattening layer. Including the input layer, the neural network has a total of 13 layers used to classify every map in 11 different categories. There is only one category representing a map with one or more ARs. To train the ConvNet we use data augmentation techniques \citep{Bishop:2006,Goodfellowetal:2016,AloysiusGeetha:2017} to increase the size of the test data. We achieve $\approx 97$~\% and $\approx 98$~\% accuracy for testing and validation, respectively. The method reliability is increased by comparing SST maps with 17~GHz images from the NoRH  what should confirm the presence of the AR.  Maps with detected ARs are \textit{binarized}: a black (1) or white (0) pixel means it does or does not belong to an AR. The AR is extracted from the map using the Canny algorithm \citep{Canny:1986}, and its physical properties: position, size and maximum excess brightness temperature are  automatically included in a data base for later analysis.  More than 16,000 maps between March 2001 and December 2017 are analyzed. Observing times are restricted to the Sun meridian transit (14:00 - 16:00 UTC) to reduce the atmospheric interference. Details of the methodology can be found in \cite{GarciaPereiraetal:2020}, along with preliminary statistical analysis.\\

In this work we concentrate in the AR excess brightness temperature \dtb\ with respect to the quiet Sun, i.e. the maximum temperature of the AR with respect to the quiet Sun. Since SST has beams with sizes of the order of a typical AR size, we do not have spatial resolution to discriminate different zones of the AR, and \dtb\ represents the average over the whole region weighted by the telescope beams.  Table \ref{tbl:AR} summarizes the observations. For every year and month, a mark \lo\ and/or \hi\ indicates that ARs are detected at 212 or 405~GHz; at the end of each line we show the total number of detected ARs during the year and the yearly median of  \dtb, the super- and underscript are the differences to the third and first quartiles of the sample, respectively. The absence of marks means either there are no observations or that no AR is detected. During some years (2003, 2004 and 2008) technical problems temporarily stopped or reduced observations. On the other hand, it can be seen that there are more ARs detected at 212 GHz than at 405~GHz, an expected result since the opacity is much higher at 405 GHz. Another remark from the table is that from mid May to mid October (Winter time) when the atmospheric humidity is lower we detect more ARs.\\

\section{Results}

\begin{table}
  \begin{center}
  \begin{tabular}{|l|c|c|c|c|c|c|c|c|c|c|c|c|c|c|c|c|}
    \hline
     Year& \multicolumn{12}{c|}{Month} & \multicolumn{2}{c|}{Total} & \multicolumn{2}{c|}{$\Delta T_\mathrm{b}$ [K]} \\
         & 1      & 2      & 3      & 4      & 5      & 6      & 7      & 8      & 9      & 10      & 11      & 12      & 212 & 405 & 212 & 405 \\
    \hline
    2001 &        &        & \lo\hi & \lo\hi & \lo\hi & \lo\hi & \lo\hi & \lo\hi & \lo\hi & \lo\hi  & \lo\hi  &         & 164 & 98  &  $ 611^{ 183}_{ 216}$ & $ 412^{1197}_{ 116}$ \\
    \hline
    2002 & \lo\hi & \lo    &        &        &        & \lo\hi & \lo\hi & \lo\hi &  \hi   & \lo\hi  & \lo\hi  & \lo\hi  & 151 & 78 & $ 362^{  85}_{  94}$ & $ 383^{  78}_{  99}$ \\
    \hline
    2003 &        &        &        &        & \lo\hi & \lo\hi &        &        &        &         &         &         &  35 & 25 & $ 305^{  43}_{  46}$ & $ 304^{  43}_{  54}$ \\
    \hline
    2004 &        &        &        &        &        &        & \lo\hi &        &        &         & \lo\hi  & \lo     &  12 & 10 & $ 298^{ 143}_{  84}$ & $ 204^{ 277}_{  38}$ \\
    \hline
    2005 &        &        &        &        & \lo\hi & \lo\hi & \lo\hi & \lo    &        &         &         &         &  92 & 86 & $ 275^{  74}_{  59}$ & $ 312^{  95}_{  83}$ \\
    \hline
    2006 &        &        &        &        &        & \lo\hi & \lo\hi & \lo\hi &        &         & \lo\hi  & \lo     &  47 & 27 & $ 246^{  87}_{  61}$ & $ 257^{  77}_{  41}$ \\
    \hline
    2007 &        &        &        &        & \lo\hi & \lo\hi & \lo\hi & \lo\hi &        &         & \lo\hi  & \lo     &  85 & 51 & $ 208^{ 163}_{  69}$ & $ 176^{ 131}_{  42}$ \\
    \hline
    2008 &        & \lo\hi & \lo    &        &        &        &        &        &        &         & \lo\hi  & \lo\hi  &  10 &  7 & $ 175^{ 187}_{  86}$ & $ 325^{ 932}_{ 173}$ \\
    \hline
    2009 &        &        & \lo\hi & \lo\hi &        & \lo\hi & \lo\hi & \lo\hi & \lo\hi & \lo\hi  & \lo\hi  & \lo\hi  &  99 & 74 & $ 133^{ 231}_{  65}$ & $ 168^{ 132}_{  60}$ \\
    \hline
    2010 & \lo    & \lo\hi & \lo\hi & \lo\hi & \lo\hi & \lo\hi & \lo\hi & \lo\hi & \lo\hi & \lo\hi  & \lo\hi  & \lo\hi  & 121 & 96 & $ 229^{ 135}_{  62}$ & $ 202^{  90}_{  53}$ \\
    \hline
    2011 &        & \lo    & \lo\hi & \lo\hi & \lo\hi & \lo\hi & \lo\hi & \lo\hi & \lo\hi & \lo\hi  & \lo\hi  & \lo\hi  & 129 & 78 & $ 317^{  95}_{  78}$ & $ 266^{  71}_{  91}$ \\
    \hline
    2012 & \lo    &        & \lo\hi & \lo\hi & \lo\hi & \lo\hi & \lo\hi & \lo\hi & \lo\hi & \lo\hi  & \lo     &         & 116 & 68 & $ 309^{  96}_{  66}$ & $ 310^{  63}_{  63}$ \\
    \hline
    2013 &        & \lo    & \lo\hi & \lo\hi & \lo    & \lo    & \lo    & \lo    & \lo    & \lo     & \lo     & \lo\hi  & 126 &  5 & $ 310^{  67}_{  65}$ & $ 386^{  70}_{ 105}$\\
    \hline
    2014 & \lo\hi & \lo    & \lo\hi & \lo\hi & \lo\hi & \lo\hi & \lo\hi & \lo\hi & \lo\hi & \lo\hi  & \lo\hi  &         &  62 & 33 & $ 364^{  77}_{  92}$ & $ 322^{  79}_{  98}$\\
    \hline
    2015 & \lo    &        & \lo\hi & \lo\hi & \lo\hi & \lo\hi & \lo\hi & \lo    & \lo\hi & \lo     & \lo     &         & 136 & 56 & $ 320^{ 109}_{  82}$ & $ 241^{ 107}_{  51}$ \\
    \hline
    2016 &        &        & \lo\hi & \lo\hi &        &        & \lo\hi & \lo\hi &        & \lo\hi  & \lo\hi  &         &  34 &  7 & $ 256^{  73}_{  37}$ & $ 299^{  21}_{ 105}$\\
    \hline
    2017 &        &       &         &        &        & \lo\hi & \lo\hi & \lo\hi & \lo    & \lo\hi & \lo\hi  & \lo      &  28 & 15 & $ 237^{  68}_{  73}$ & $ 229^{ 248}_{  41}$\\
    \hline
    \multicolumn{13}{|r|}{\textbf{Total Number of ARs detected}}                                                         &1473 &826  &\multicolumn{2}{c|}{} \\
    \hline
  \end{tabular}
  \caption{Summary of AR observations. The symbol \lo\ (\hi) represents one or more ARs detected at 212 (405) GHz during the month. No symbol means either no detection or no observation. The columns labeled Total show the total number of ARs detected during the year, and the last two columns give median \dtb\ for the year with the $\pm$ uncertainties defined as the difference to the third quartile (superscript) and first quartile (subscript) of the sample.}
    \label{tbl:AR}
  \end{center}
\end{table}

\subsection{Excess Brightness Temperature Time Evolution}

\begin{figure} 
  \centerline{\includegraphics[width=0.85\textwidth]{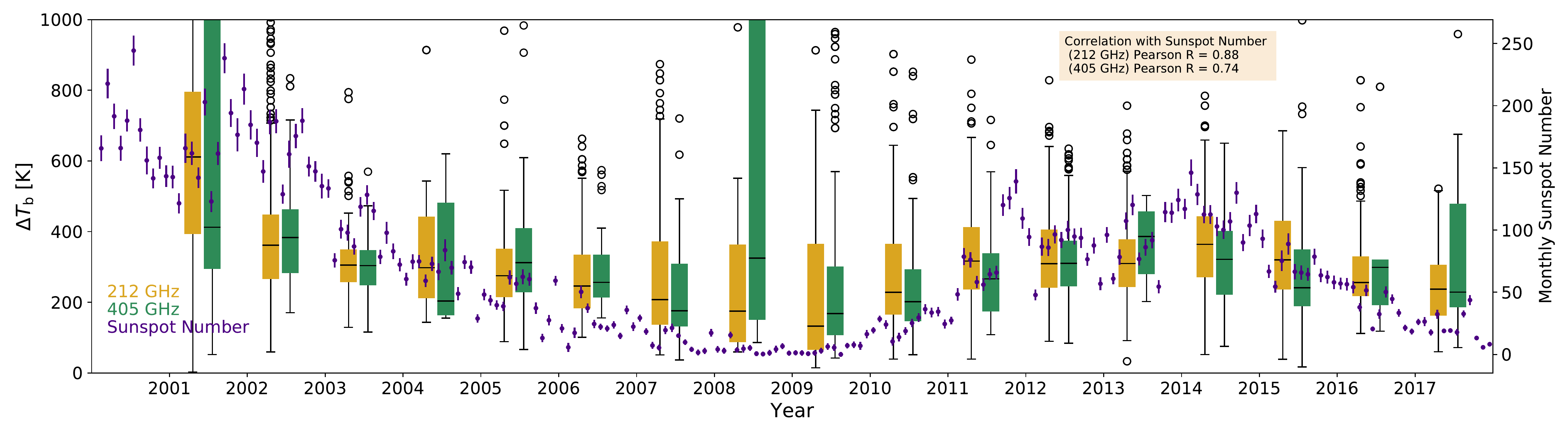}}
  \caption{Boxplot time-series for \dtb\ at 212 (golden) and 405 GHz (green). Every box represents data for a whole year: the horizontal black line inside the box is the year median. Purple dots are the monthly mean sunspot number, with their error bars defined as 1 $\sigma$. The open circles represent the outliers of the distribution. }
  \label{fig:TbEvolution}
\end{figure}

Figure \ref{fig:TbEvolution} shows the evolution of the AR excess brightness temperature $\Delta T_\mathrm{b}$ for the two SST observing frequencies in the form of \textit{boxplots}. Every box represents one year of observations, the central, horizontal black line is the median; the lower (upper) limit of the box is the first (third) quartile, and the vertical line extends from the minimum to the maximum; the open circles represent the outliers of the distribution. Golden boxes represent results for 212 GHz while green boxes for 405 GHz. The purple dots are the monthly mean values of the sunspot number (SSN) obtained from the Sunspot Index and Longterm Solar Observations (SILSO) which is an activity of the Solar Influences Data Analysis Center (SDIC) of the Royal Observatory of Belgium. The error bars correspond to 1~$\sigma$ for the month. We can see that \dtb\ at the two frequencies follow the solar cycle represented by the SSN. The overall cross correlation coefficients between the annual median \dtb\ and the annual mean sunspot number are $R=0.88$ for 212 GHz and $R=0.74$ for 405 GHz; but if we restrict to the years 2009 and 2017, $R=0.93$ and $0.81$ for 212 and 405 GHz, respectively. One reason why $R$ increases after 2009 is that the antenna was in commissioning during almost the first decade after installation in 1999. During 2008, in particular, we detected ARs during three months only, resulting in a small sample that produces a significant increase in the uncertainty (see the boxplot for 405 GHz in Figure \ref{fig:TbEvolution}). On the contrary, after 2009, we observe that boxplots are smaller, an indication of a narrow data dispersion.

\begin{figure} 
  \centerline{\includegraphics[width=0.8\textwidth]{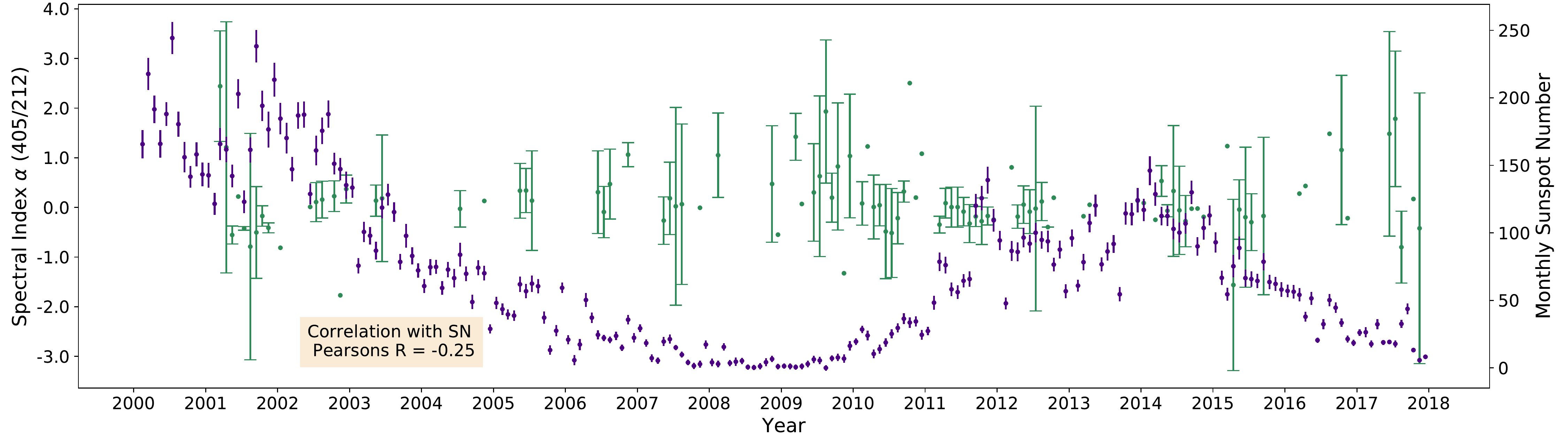}}
  \caption{Green dots are the monthly median of the spectral index
    $\alpha$, with error bars representing 1
    $\sigma$. No error bar means only one measurement. Purple circles
    are the monthly sunspot number.}
  \label{fig:alphaEvolution}
\end{figure}

\subsection{Spectral Index}

\begin{figure} 
  \centerline{\includegraphics[height=0.3\textwidth]{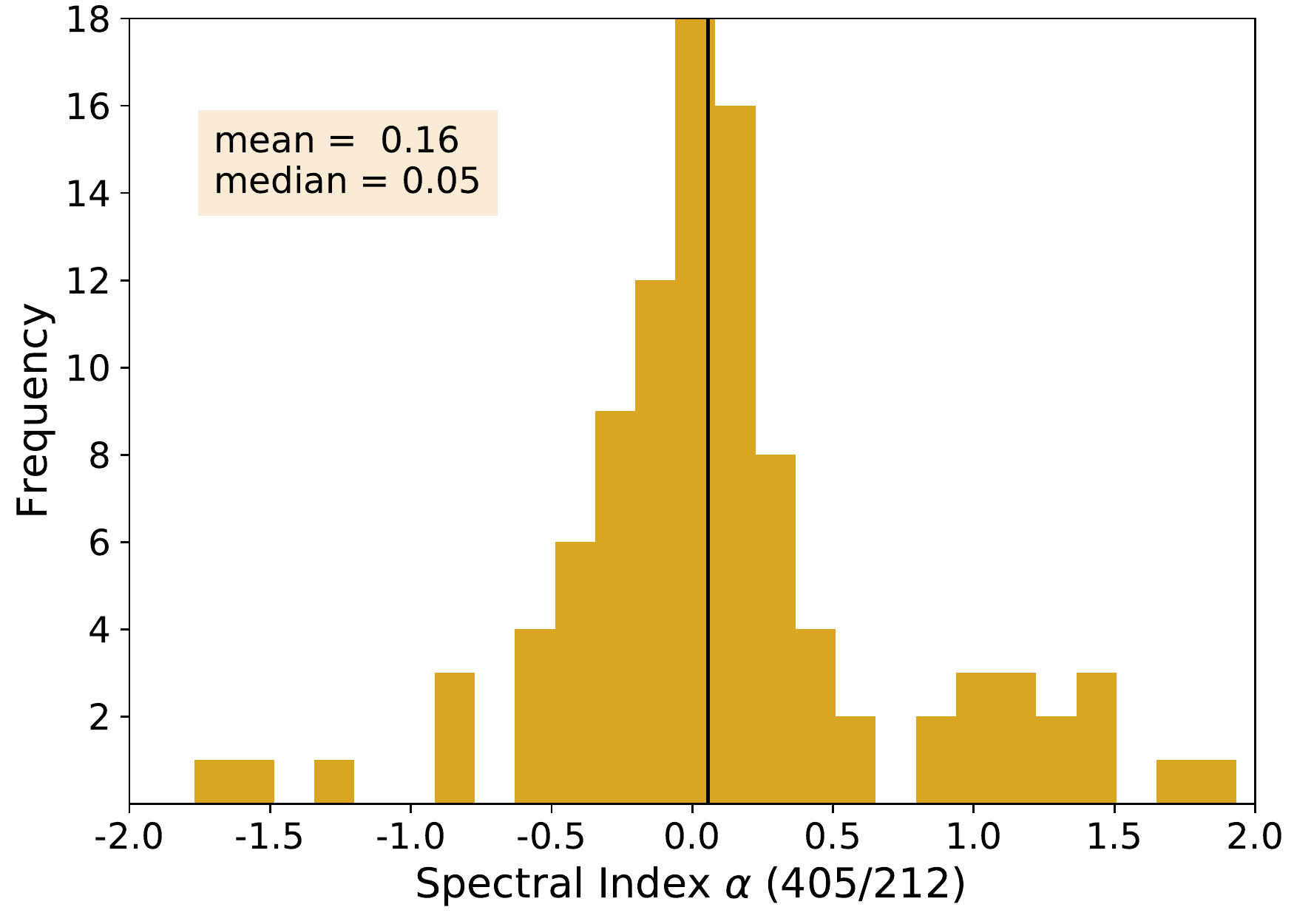}\hspace{1cm}
  \includegraphics[height=0.3\textwidth]{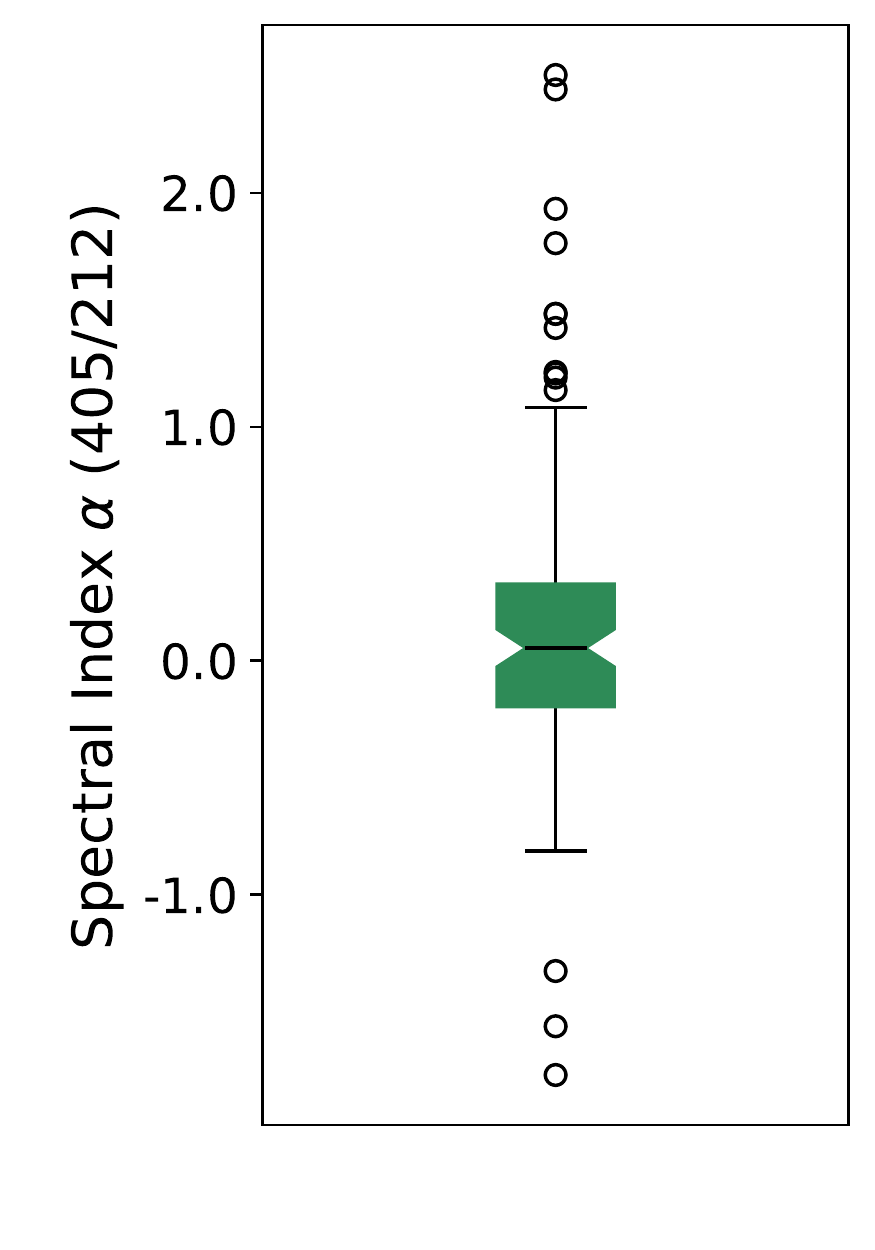}}
  \caption{Left. Frequency distribution of the spectral index
    $\alpha$, the vertical black line represents the median
    $\alpha_\mathrm{med}=0.05$. Right. A boxplot representing the sample.}
  \label{fig:Alpha}
\end{figure}

From \dtb\ we can derive the spectral index $\alpha$ defined as
\begin{equation}
  \alpha = \frac{\log(\Delta T_{\mathrm{b}_{405}})-\log(\Delta T_{\mathrm{b}_{212}})}{\log(405)-\log(212) } \ ,
  \label{eq:alpha}
\end{equation}
where $\Delta T_{\mathrm{b}_{405}}$ and $\Delta T_{\mathrm{b}_{212}}$ are the excess brightness temperatures at 405 and 212 GHz, respectively. The temporal evolution of $\alpha$ is shown in Figure \ref{fig:alphaEvolution}, green dots are the monthly mean while the uncertainty bars correspond to 1~$\sigma$, purple dots are the sunspot number as in Figure \ref{fig:TbEvolution}.  In Figure \ref{fig:Alpha} we see the frequency distribution of $\alpha$, and a boxplot of the sample; both ways graphically show the rather symmetric distribution with a mean value $\langle\alpha\rangle = 0.16$ and median $\alpha_\mathrm{med} = 0.05$; implying that the sample dispersion is related to random noise. On the other hand, the overall cross correlation coefficient  $R=-0.25$ is consistent with an apparent slight anti-correlation observed in Figure \ref{fig:alphaEvolution}.\\

\section{Discussion}

Previous works with better spatial resolution have shown that at submillimeter wavelengths the sunspot umbra is cooler than the quiet Sun, while the penumbra has a similar temperature and the surrounding plage is much hotter \citep{LindseyKopp:1995,Loukitchevaetal:2017,Iwaietal:2017}. However, SST beam sizes, which are of the same order of the AR, cannot resolve these spatial variations and \dtb\ is an average over the whole region, with the penumbra and surrounding plages dominating the emission, i.e. what we are determining is the variation of the umbra + penumbra + plage temperature averaged by the beam size along the solar cycle.  \cite{ValleSilvaetal:2020} have shown that there are many ARs observed at submillimeter wavelengths without a spot, as it was previously seen by \cite{Selhorstetal:2014} at cm-wavelength. We also note that, since we use the quiet Sun as a calibrator, and we assume that the quiet Sun brightness temperature does not change with the cycle, the absolute \dtb\ may be different at different phases of the cycle if this hypothesis is not correct. However, even if the quiet Sun temperature changes during the cycle, \dtb\ variations are real.\\

Cycles 23 and 24 were double peaked with an evident gap in between them \citep[see][]{Gnevyshev:1967}, Figure \ref{fig:TbEvolution}  shows this behavior for the SSN. SST started observations in 2001, during the recovery from the Gnevyshev gap that was relatively fast and took less than a year for the SSN. Therefore, it is not possible to identify the gap in the \dtb\ time-series. In Cycle 24,  the gap between peaks occurred between 2012-2013,  which is not apparent in the  \dtb\ time-series. We cannot, however, draw any firm conclusion about this point, since SST data are limited by the atmospheric attenuation and beam sizes that reduce the detected AR peak temperature, which may create \textit{false-negatives}. For this reason we present only one year average values. We are working now in an enhanced AI method to increase the AR detectability in the SST data.\\

As it was already pointed out, \cite{Selhorstetal:2014} have shown the correlation between the ARs brightness temperature and the solar cycle at 17~GHz. This is an expected result at this frequency due to the emergence of ARs with strong magnetic fields \citep[$|B|\geqslant 2200$~G at the photosphere, ][]{Vourlidasetal:2006}. These ARs may present gyro-resonance contribution, that is able to increase their brightness temperatures from $2-3\times 10^4$~K to values as high as $T_B>10 ^5$~K, if the 17~GHz gyro-resonance third harmonic (2000~G) occurs above the transition region \citep{Shibasaki:1994,Vourlidasetal:2006}. However, within solar conditions, the same mechanism does not apply at submillimeter wavelengths. The millimeter to submillimeter emission is mostly a thermal continuum produced at chromospheric height in local thermodynamic equilibrium \citep[LTE, e.g. ][]{Loukitchevaetal:2017}. Instead of the flux density, as it was used in previous works and which depends on the brightness temperature and the source solid angle, here we only use the brightness temperature to characterize the spectrum, reducing the uncertainty of the spectral index. Indeed, after the propagation of temperature uncertainties in Equation (\ref{eq:alpha}), the $\alpha$ uncertainty is

\newcommand{\dtbh}{\Delta T_{\mathrm{b}_{405}}}
\newcommand{\dtbl}{\Delta T_{\mathrm{b}_{212}}}

\begin{eqnarray}
  \Delta^2\alpha &=& \left ( \frac{\partial \alpha}{\partial \dtbh}\right )^2 \Delta^2(\dtbh) +
                     \left ( \frac{\partial \alpha}{\partial \dtbl}\right )^2 \Delta^2(\dtbh) \nonumber \\
                 &=& \left ({\cal Q}\ \frac{ \Delta(\dtbh)}{\dtbh}\right )^2  +
                     \left ({\cal Q}\ \frac{\Delta(\dtbl)}{\dtbl}\right )^2  = {\cal Q} (\varepsilon^2_{r_{405}} + \varepsilon^2_{r_{212}}) \ ,
\label{eq:alpha-unc}
\end{eqnarray}
where $\Delta(\dtbh)$ and $\Delta(\dtbl)$ are the total uncertainties in excess brightness temperatures for 405 and 212~GHz, respectively, while $\varepsilon_{r_{405}}$ and $\varepsilon_{r_{212}}$ are their relative uncertainties and ${\cal Q} = 1/\log(405/212)$. Substituting $\varepsilon_{r_{405}}=\varepsilon_{r_{212}}\approx 0.1$ in Equation (\ref{eq:alpha-unc}) we get $\Delta\alpha \approx 0.22$, a value which agrees with Figure \ref{fig:Alpha} and reinforces our conclusion that the dispersion is produced by random noise. Therefore,  we are very cautious to draw any conclusion about the anti-correlation between $\alpha$ and the SSN given by the coefficient $R=-0.25$. The spectral index $\alpha$ obtained here represents the average source properties, encompassing plages, penumbra and umbra from one or more sunspots; however, we note that the spectrum of an AR can be well represented by a unique spectral index from 34~GHz to 405~GHz when the spatial resolution is of order 1~arcmin \citep{Silvaetal:2005,ValleSilvaetal:2020} lending support to the optically thick emission of a thermal source.\\

The temporal variation of \dtb\ has to be also interpreted in terms of an average value. The evolution of an active region is modulated by the magnetic field; it is the magnetic energy dissipated the one that should heat the plasma.   As a solar cycle progresses the dissipated magnetic  energy should increase or decrease following it, and therefore, the active region average temperature should follow the cycle.  However, to confirm this hypothesis, a statistical study relating the photospheric magnetic field and \dtb\ should be accomplished.\\

\acknowledgments

This work is based on data acquired at Complejo Astron{\'o}mico El Leoncito, operated under agreement between the Consejo Nacional de Investigaciones Cient{\'\i}ficas y T{\'e}cnicas de la Rep{\'u}blica Argentina and the National Universities of La Plata, C{\'o}rdoba and San Juan. The research leading to these results has received funding from CAPES grant 88881.310386/2018-01, FAPESP grant 2013/24155-3 and the US Air Force Office for Scientific Research (AFOSR) grant FA9550-16-1-0072 and by Mackenzie Research Funding Mackpesquisa. Brazilian Federal Agency CAPES and S{\~a}o Paulo State Research Agency FAPESP have supported AGP and JFVS with student and PD fellowships. CGGC, CLS and JPR are grateful to CNPq for support with the Productivity Research Fellowships. CHM and GDC acknowledge financial support from Argentine grants PICT 2016-0221 (ANPCyT) and UBACyT 20020170100611BA. CHM, GDC and CGGC are members of the Carrera del Investigador Cient{\'\i}fico of the Consejo Nacional de Investigaciones Cient{\'\i}ficas y T{\'e}cnicas (CONICET). This research used version 1.1.3 of the SunPy open source software package \citep{sunpy_community:2020}.

\bibliographystyle{aasjournal}

\end{document}